\documentclass[11pt]{article}
\usepackage{amsmath,fullpage,graphicx,booktabs,palatino,epsfig,fancyhdr,color,amssymb,relsize,setspace}

\renewcommand{\headrule}{{\hrule width\headwidth height\noheadrule}}

\newcommand{\appsection}[1]{\let\oldthesection\thesection
\renewcommand{\thesection}{Appendix \oldthesection}
\section{#1}\let\thesection\oldthesection}

\begin{document}

\begin{figure*}[h]
\begin{center}
\vspace{-1.0cm}
\includegraphics[height=4.0cm]{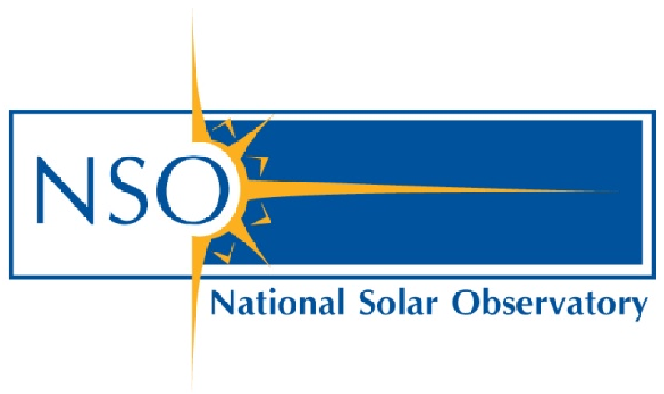}
\end{center}
\end{figure*}

\begin{center}

\vspace{2.0cm}

\begin{LARGE}
\textbf{Ca II K 1-\AA~ Emission Index Composites}
\end{LARGE}

\vspace{3.5cm}

\begin{large}

Luca Bertello, Andrew R. Marble, \& Alexei A. Pevtsov

\vspace{0.5cm}

National Solar Observatory

\vspace{3.5cm}

\today

\vspace{4.5cm}

\hrule

\vspace{1.0cm}

Technical Report No. \textbf{NSO/NISP-2017-001}

\end{large}

\end{center}

\clearpage
\renewcommand{\headrule}{{\hrule width\headwidth 
height\headrulewidth\vskip0.5cm}}
\fancyhead{}
\fancyhead[L]{Ca II K 1-\AA~ Emission Index Composites}
\fancyhead[R]{\thepage}
\fancyfoot{}
\thispagestyle{fancy}

\vspace*{-0.40cm}

\begin{center}
\begin{Large}
\bf{Abstract}
\end{Large}
\end{center}

\begin{quote}

We describe here a procedure to combine measurements in the 393.37 nm Ca II K spectral line 
taken at different observatories. Measurements from 
the National Solar Observatory (NSO)
Integrated Sunlight Spectrometer (ISS) on the Synoptic Optical Long-term Investigations of the Sun (SOLIS)
telescope, the NSO/Sac Peak Ca II K-Line Monitoring Program, and Ca II K filtergrams from Kodaikanal Solar Observatory 
(KKL) are merged together to create a pair of composites of the Ca II K 1-\AA~ emission index. 
These composites are publicly available from the SOLIS website at 
http://solis.nso.edu/0/iss/.
\end{quote}

\section{Introduction}

Measurements of the ionized Ca II K-line are one of the major resources for long-term studies of solar and stellar activity. 
They also play a critical role in many studies related to solar irradiance variability, particularly as a ground-based proxy to 
model the solar ultraviolet flux variation that may influence the Earth's climate. 
Full disk images of the Sun in Ca II K have been available from various observatories for more than 100 years, 
while synoptic Sun-as-a-star observations in Ca II K began in the early 1970s. 
In 1976, observations of the disk-integrated solar Ca II K-line began at the National Solar Observatory (NSO)/ Sacramento Peak
(Sac Peak). 
After October 1, 2015, the NSO Ca II K-line series continued using the
Integrated Sunlight Spectrometer (ISS) on the Synoptic Optical Long-term Investigations of the Sun (SOLIS) facility. 
The ISS has been in operation since late 2006  at NSO/Kitt Peak. In July 2014, SOLIS was relocated to 
a temporary facility in Tucson. In 2017, it will be moved to a new permanent location, which remains to be determined. 
The ISS
obtains high (R $\sim$ 300,000) spectral resolution daily observations of the Sun as a star in
nine different spectral bands that reside in a broad range of wavelengths (350 nm - 1100 nm), 
including the Ca II K band centered at 393.37 nm. 
Several K-line parameters, including the emission index and various measures of asymmetry, are computed from the calibrated 
Ca II K-line 
profiles. One of these parameters, the 1-\AA~ Ca II K emission index, is defined as the equivalent width of a 1 angstrom band 
centered on the K-line core.
Although the ISS was designed as a future replacement for Sac Peak K-line monitor, there are some differences in
both the
K-line profile observations and the parameter derivation. 
For example, ISS equivalent widths reflect a novel approach based 
on the Fourier components of the line profile (Bertello et al. 2011), and ISS intensity calibration employs a more sophisticated 
approach than that of the Sac Peak K-line monitor (Pevtsov, Bertello, and Marble, 2014). 
To ensure successful inter-calibration, Sac Peak K-line monitor and SOLIS/ISS co-observed 
over a 9-year period including different levels of solar activity. This report discusses the inter-calibration of time series 
from these two instruments to form a continuous time series of sun-as-a-star K-line observations.   

\begin{figure*}[h]
\begin{center}
\includegraphics[width=\linewidth]{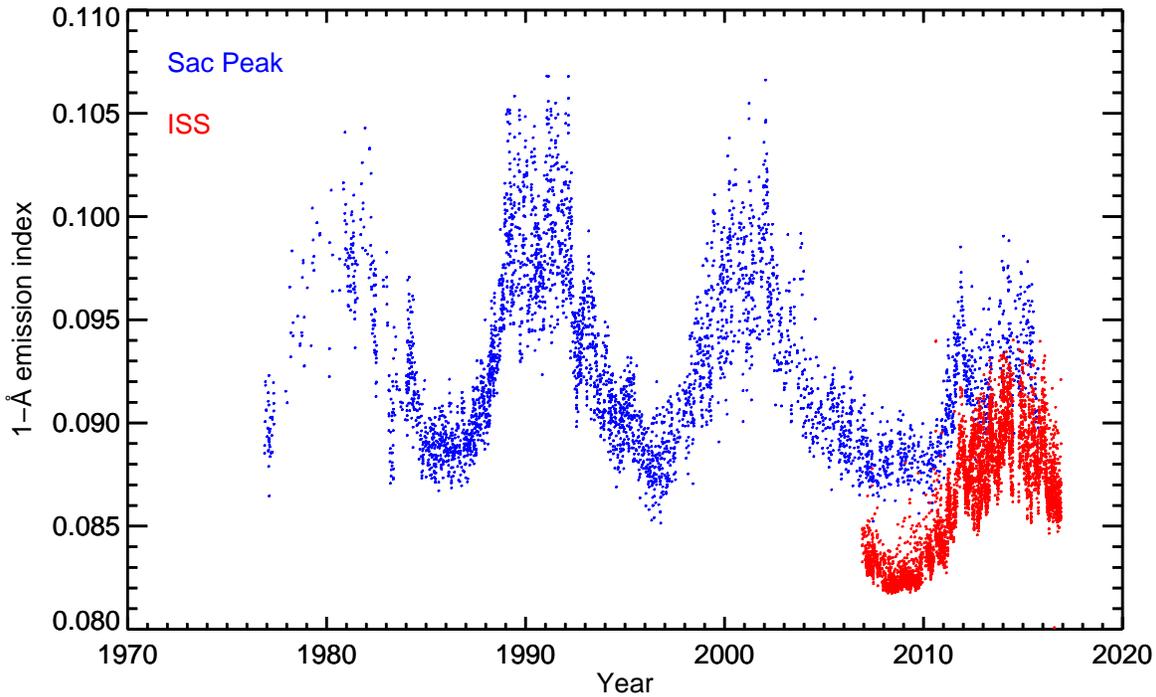}
\caption{Comparison between independently calibrated daily Sac Peak (blue) and ISS (red) time series
of the Ca II K 1-\AA~ emission index.}
\label{raw}
\end{center}
\end{figure*}

Figure \ref{raw} shows both ISS and Sac Peak 1-\AA~ emission index time series, independently calibrated. There
is an evident discrepancy between the two sets of values.

\section{Correction to the Sac Peak Measurements}

\begin{figure*}[h]
\begin{center}
\includegraphics[width=\linewidth]{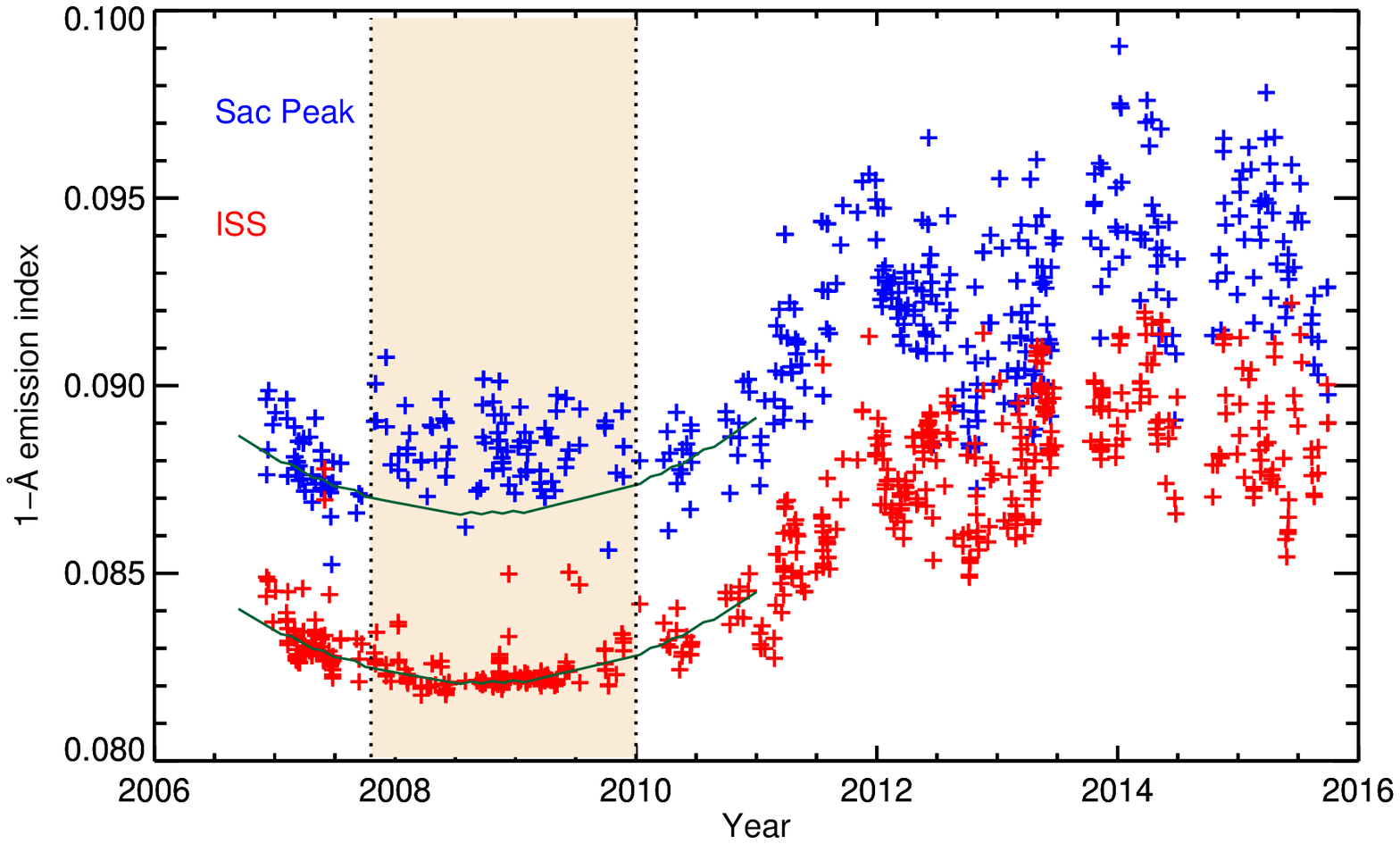}
\caption{Same day measurements taken at the two sites. Each time series has been independently calibrated, resulting
in a slightly different scaling between the two data sets. The green curve shows the expected behavior of the 1-\AA~ emission
index during the extended minimum of solar cycle 23/24 according to the ISS data. During that time, highlighted by the
'antique white' colored area, Sac Peak data show some unresolved calibration issues.}
\label{initial}
\end{center}
\end{figure*}

Figure \ref{initial} shows the daily values of the 1-\AA~ emission index derived from measurements taken
on the same day, around local noon, 
at both sites. In addition to a different scaling, the
Sac Peak values seem to have some additional unresolved calibration issues during the extended minimum of cycle 23/24. 
From the
second half of 2007 to the end of 2009, those values seem to deviate from the trend of the overall time series, as
highlighted in the plot by
the filled area colored in 'antique white'. In particular, the 
Sac Peak 1-\AA~ emission index seems to be slightly overestimated
during that time. 

Assuming the same temporal trend for both data sets, we first fit a quadratic function to the ISS values
over the time interval [2007,2011]. 
In Figure \ref{initial},  
this function is shown as a green curve. The ISS data are nicely distributed
around this function, with very few outliers. 
The same function is then multiplied by 1.055 to model the Sac Peak data over the same
time interval. The
1.055 scaling is determined empirically to provide the best fit with the ISS data. 
It is quite evident from a simple visual
inspection of Figure \ref{initial}, that the Sac Peak data between 2007.8 and the end of 2009 appear to systematically
overestimate the emission index value. Although the exact reason for this discrepancy has not been found, it is
possible that some instrumental effects have not been accounted for in the original calibration of these data.

\begin{figure*}[h]
\begin{center}
\includegraphics[width=0.49\linewidth]{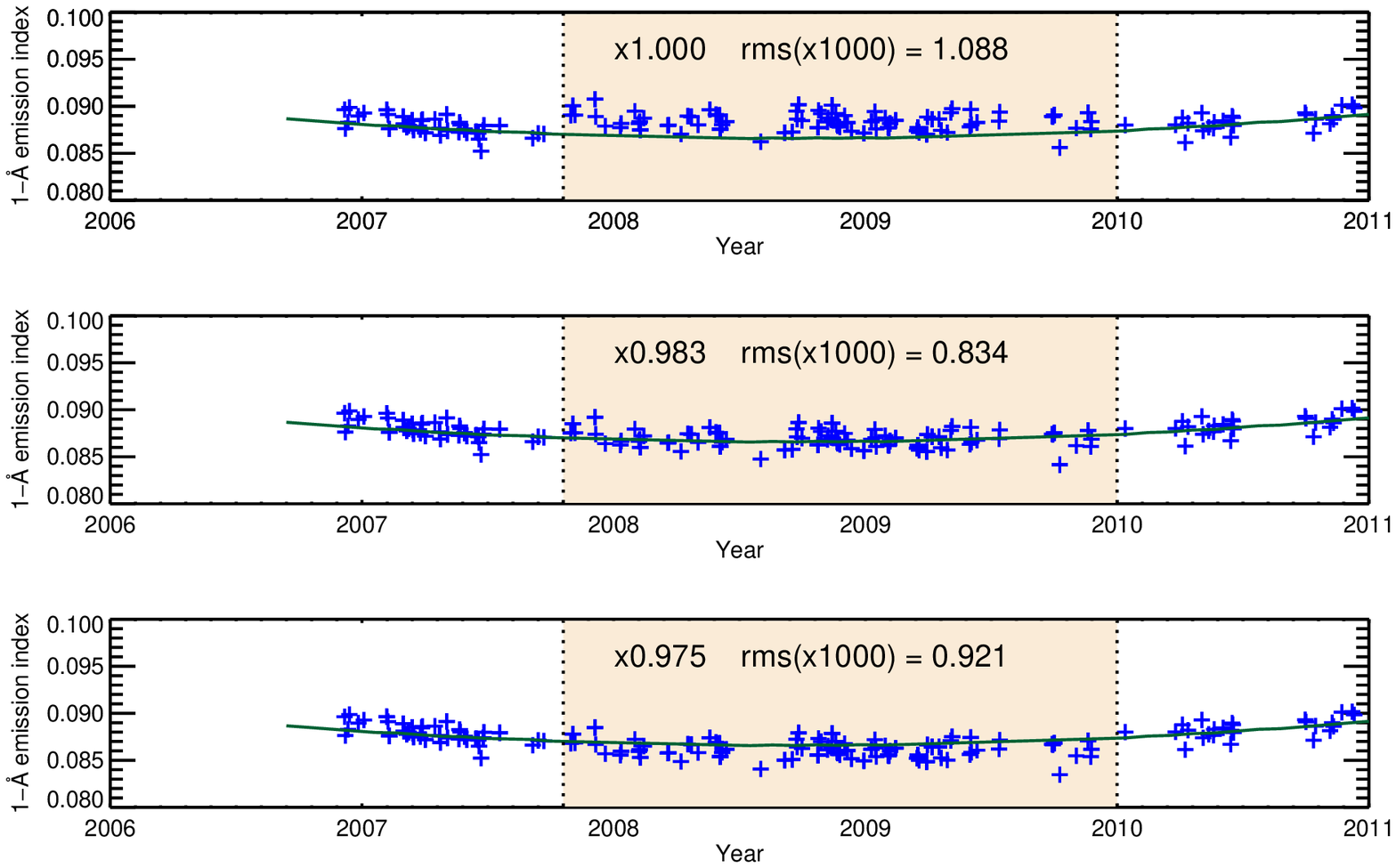}
\includegraphics[width=0.49\linewidth]{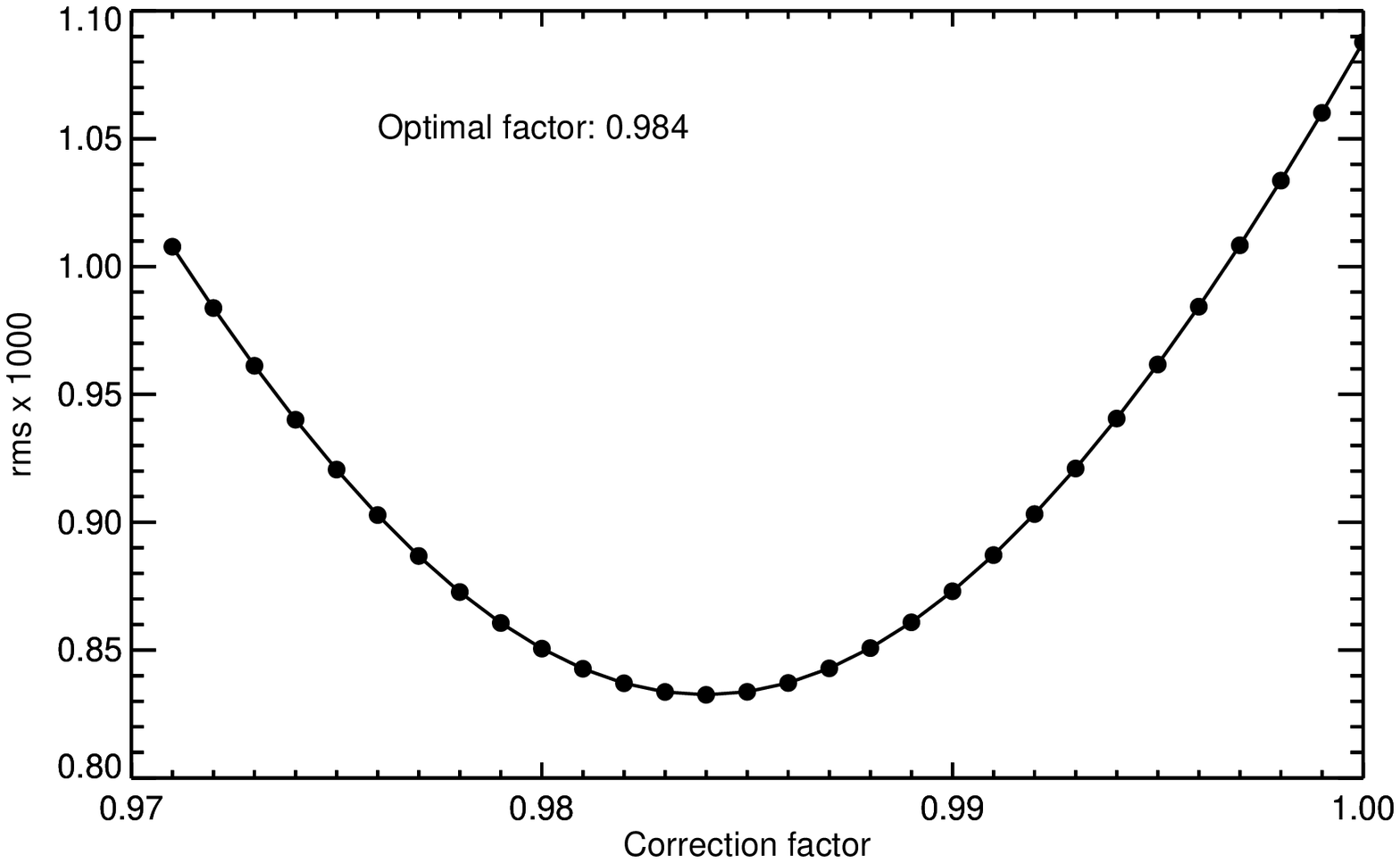}
\caption{Determination of the adjusting factor required to correct the Sac Peak data between 2007.8 and 2010.0, 
shown between the light-brown area. Left plot shows
the results after multiplying those data by three different scaling factors. The right plot illustrates how different
correction factors affect the rms of the difference between the data and the model. 
The correction factor minimizing the rms is selected to be the optimal factor.}
\label{factor}
\end{center}
\end{figure*}

To correct those
measurements, we multiply them by different scaling factors and determine the one that
minimize the rms of the difference between the adjusted data and the model (green curve in Figure \ref{initial}).
The left panel
of Figure \ref{factor} shows the results of using three different scaling factors. The middle plot
shows an almost ideal value. The right panel illustrates in detail the determination of the optimal factor. Using a
set of possible values between 0.97 and 1.00, the rms is calculated for each correction factor. The location corresponding
to the minimum value in this curve is selected to be the optimal factor, which turned out to be 0.984.

\begin{figure*}[ht]
\begin{center}
\includegraphics[width=\linewidth]{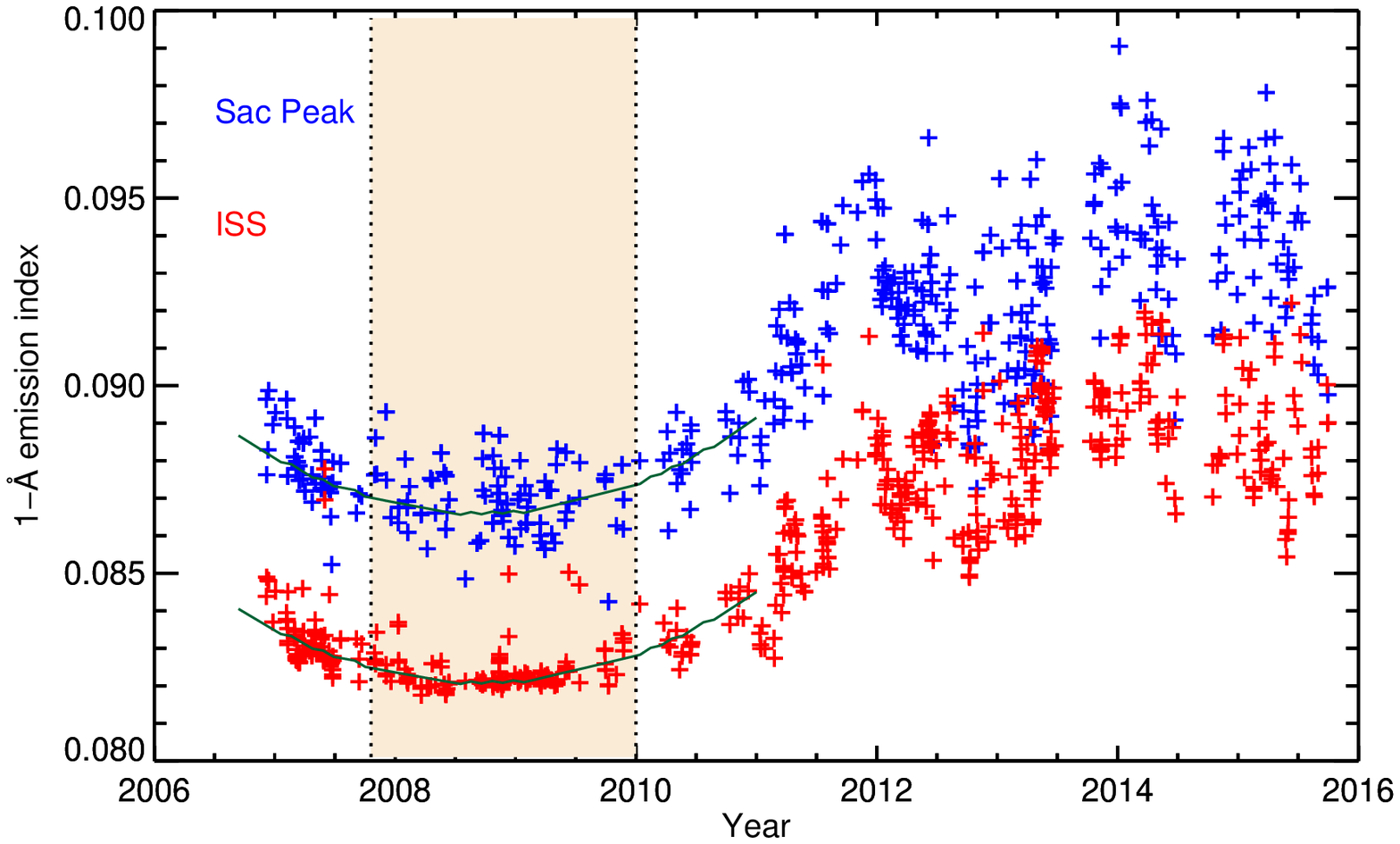}
\caption{Same as Figure \ref{initial}, after the Sac Peak data between 2007.8 and 2010.0 have been multiplied by
0.984.}
\label{final}
\end{center}
\end{figure*}

Figure \ref{final} shows the corrected Sac Peak series, after the Sac Peak data between 2007.8 and 2010.0 have been multiplied
by 0.984. Although the scatter of values is quite large compared to the corresponding ISS values, the trends of
the two time series appear to be consistent.

\section{Sac Peak and ISS Composite}

\begin{figure*}[ht]
\begin{center}
\includegraphics[width=\linewidth]{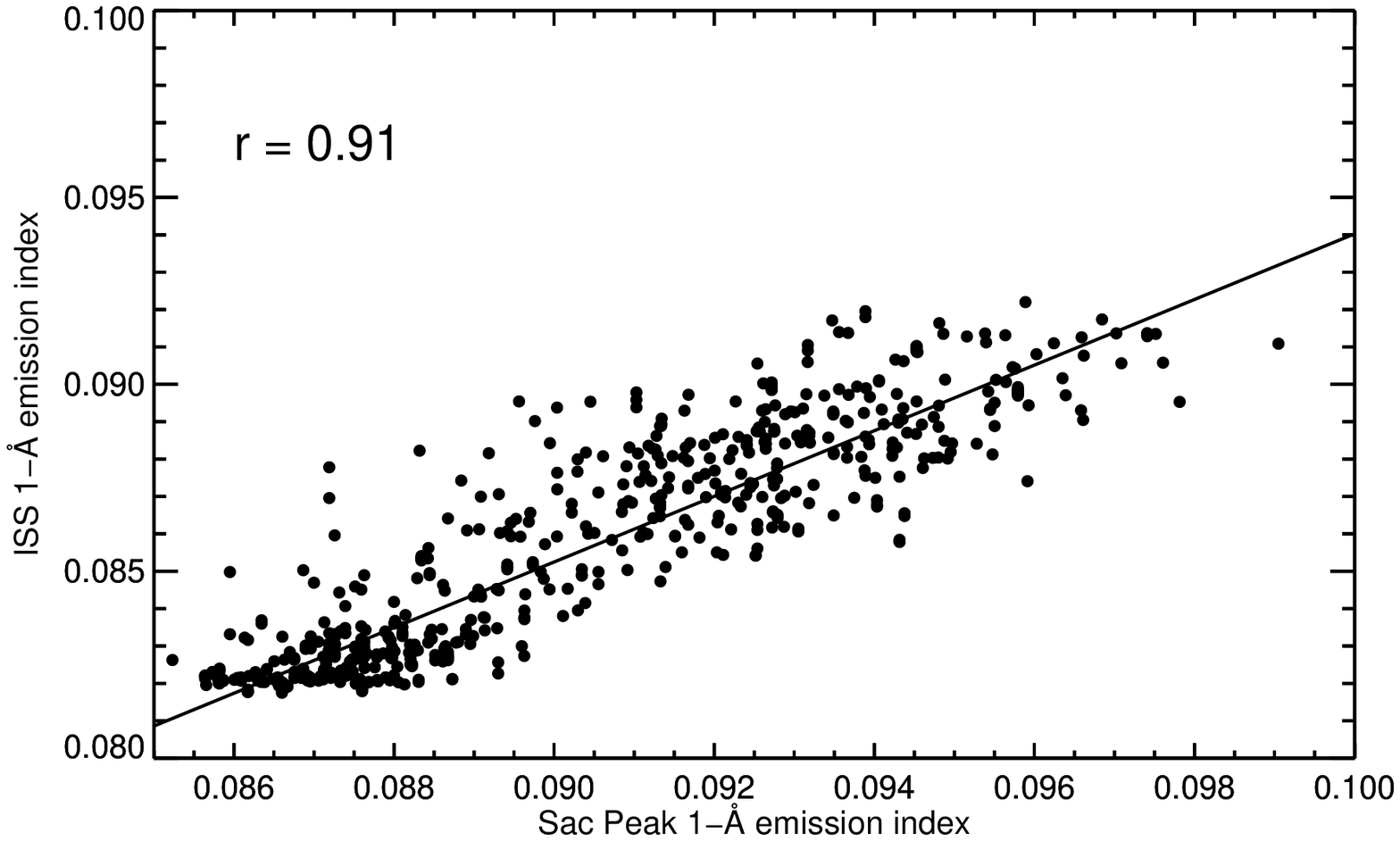}
\caption{Scatter plot of ISS and Sac Peak same day measurements around noon local time. The
high correlation between the two quantities is indicated by the correlation coefficient $r$. The coefficients
from the linear regression
are used to adjust the Sac Peak data into the ISS scale (see text).}
\label{scatter}
\end{center}
\end{figure*}

\begin{figure*}[ht]
\begin{center}
\includegraphics[width=\linewidth]{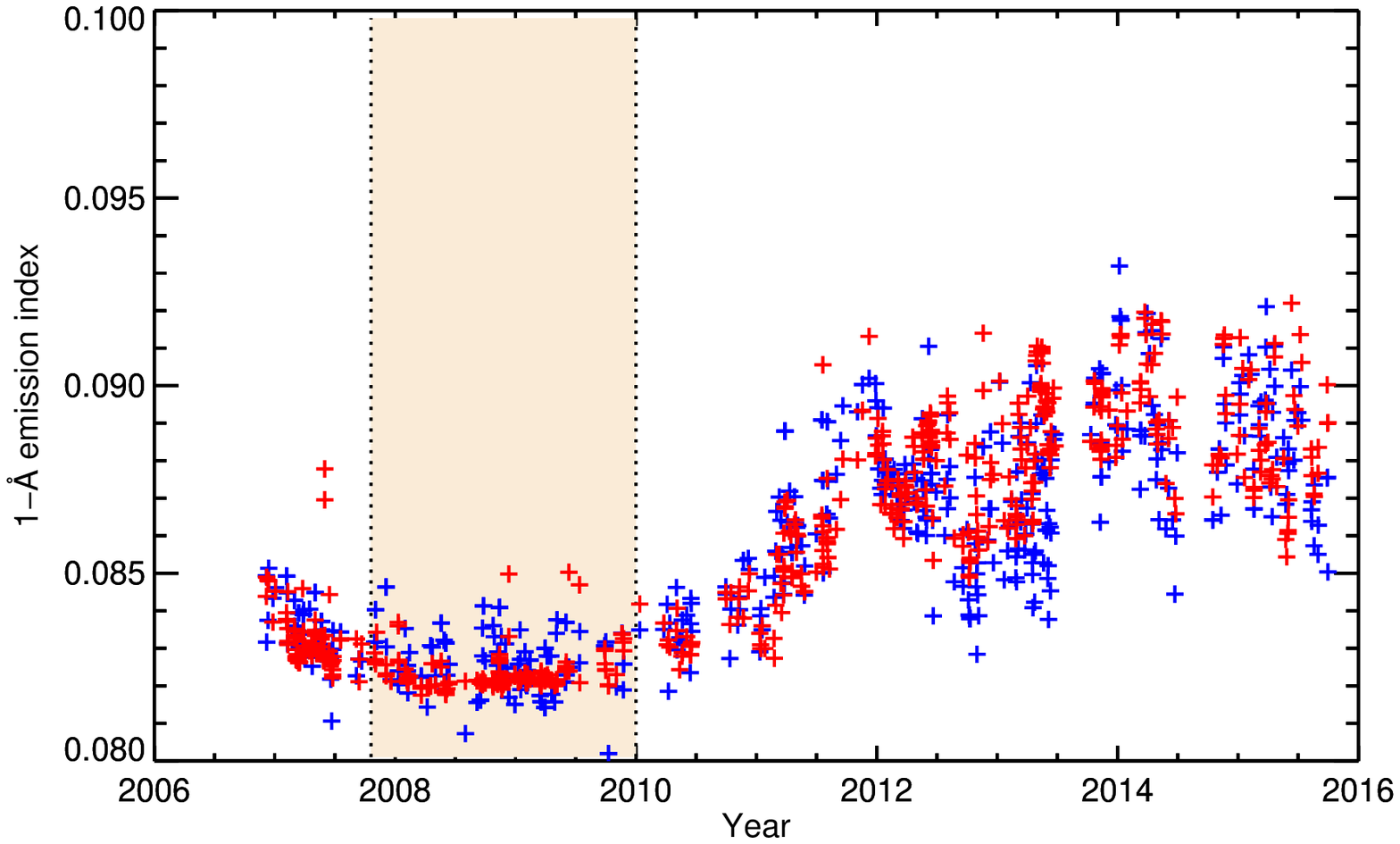}
\caption{Same as Figure \ref{final}, after the Sac Peak values have been converted into the ISS scale according
to the equation given in the text.}
\label{daily_final}
\end{center}
\end{figure*}

To combine Sac Peak K-line monitor and ISS data into a single time series, we use ISS as a reference. 
At the minimum of cycle 23, the 1-\AA~ emission index from ISS is lower in amplitude as compared to Sac Peak data.
To scale the Sac Peak values into the same scale as ISS, a simple
linear regression between the two sets of quasi-simultaneous daily values is performed. Such a linear regression
is shown in Figure \ref{scatter}. The correlation coefficient is 0.91, with a statistical significance much greater
than 99.99\%. The equation of this linear regression provides the necessary coefficients to convert
the Sac Peak data into the same scale as ISS. This transformation is given by:

\medskip
SP$_{\rm ISS}$ = (0.0062 $\pm$ 0.0015) + (0.8781 $\pm$ 0.0163) $\times$ SP$_{\rm orig}$,
\medskip

\noindent
where SP$_{\rm orig}$ and SP$_{\rm ISS}$ are the original and rescaled Sac Peak 1-\AA~ emission index values, respectively.
Figure \ref{daily_final} shows the rescaled Sac Peak 1-\AA~ emission index after the above transformation was applied.
The two data sets appear now to be fully consistent. 

\begin{figure*}[ht]
\begin{center}
\includegraphics[width=\linewidth]{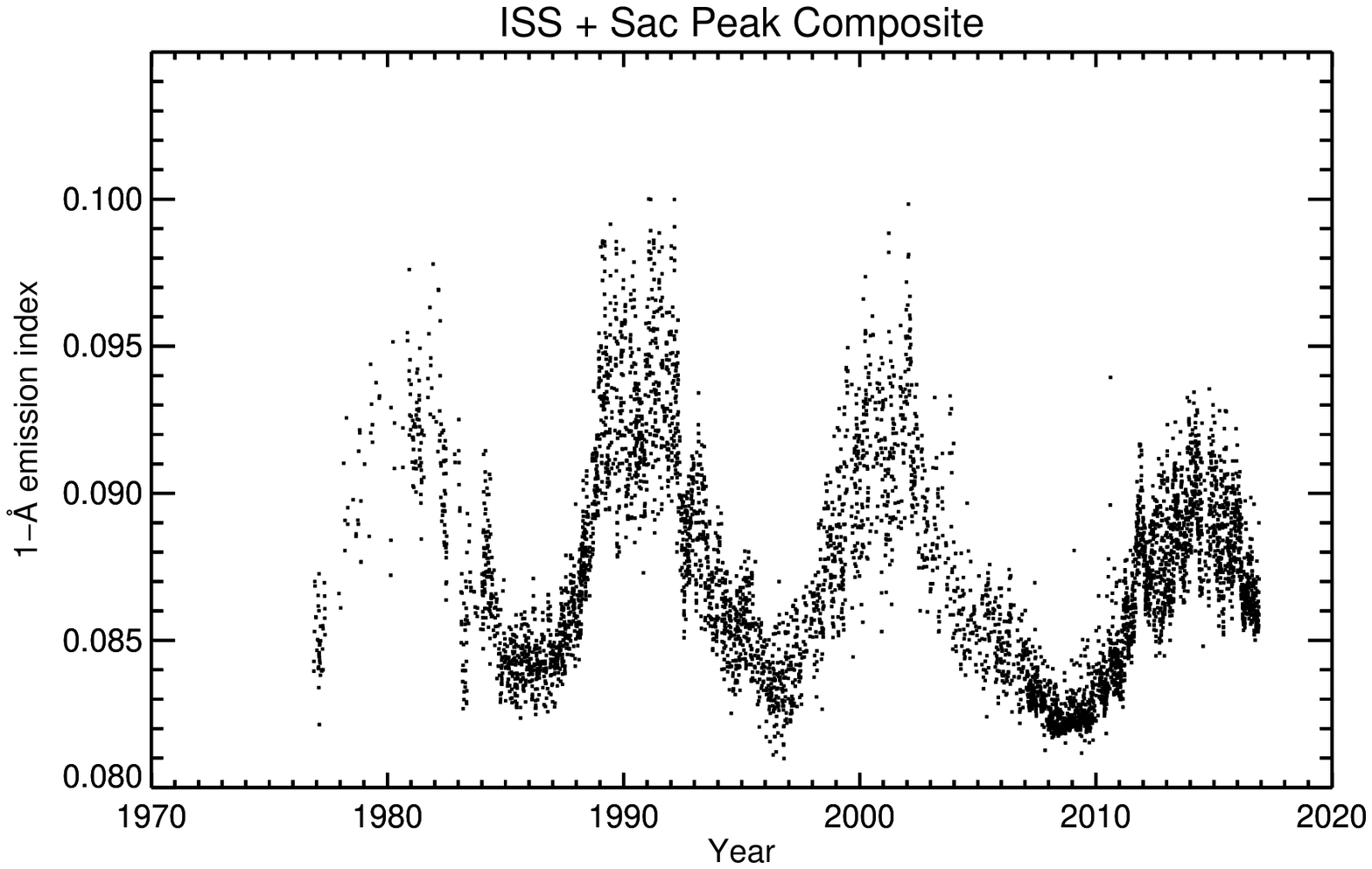}
\caption{Daily measurements of the NSO 1-\AA~ emission index composite computed using data from observations taken at
Sacramento Peak and with the ISS instrument.}
\label{merged}
\end{center}
\end{figure*}

The Sac Peak and ISS composite is created using only a single observation (typically the one closest to local noon) 
from a single site per day.  For days with both ISS and Sac Peak observations, those from the ISS are used.
This composite (hereafter NSO) covers about 40 years of NSO integrated sunlight Ca II K observations, and 
is shown in Figure \ref{merged}. 

\section{Kodaikanal and NSO Composite}

\begin{figure*}[ht]
\begin{center}
\includegraphics[width=\linewidth]{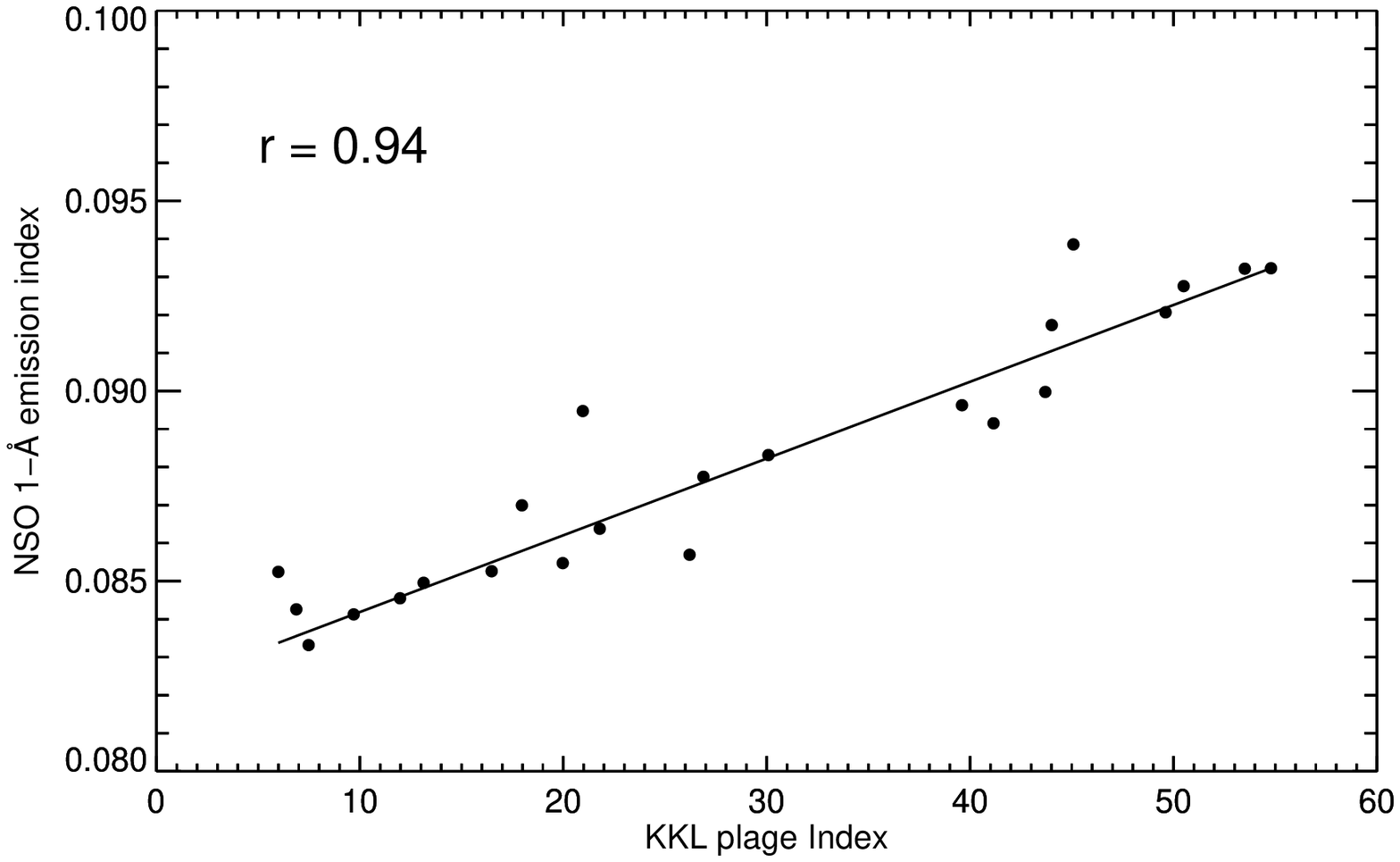}
\caption{Scatter plot of NSO and Kodaikanal (KKL) yearly average values. The regression line provides
the coefficients for the transformation of the KKL plage index into a 1-\AA~ emission index consistent with
the NSO scale. The correlation coefficient $r$ is statistically significant at a 99.9\% level.}
\label{nso_kkl}
\end{center}
\end{figure*}

\begin{figure*}[ht]
\begin{center}
\includegraphics[width=\linewidth]{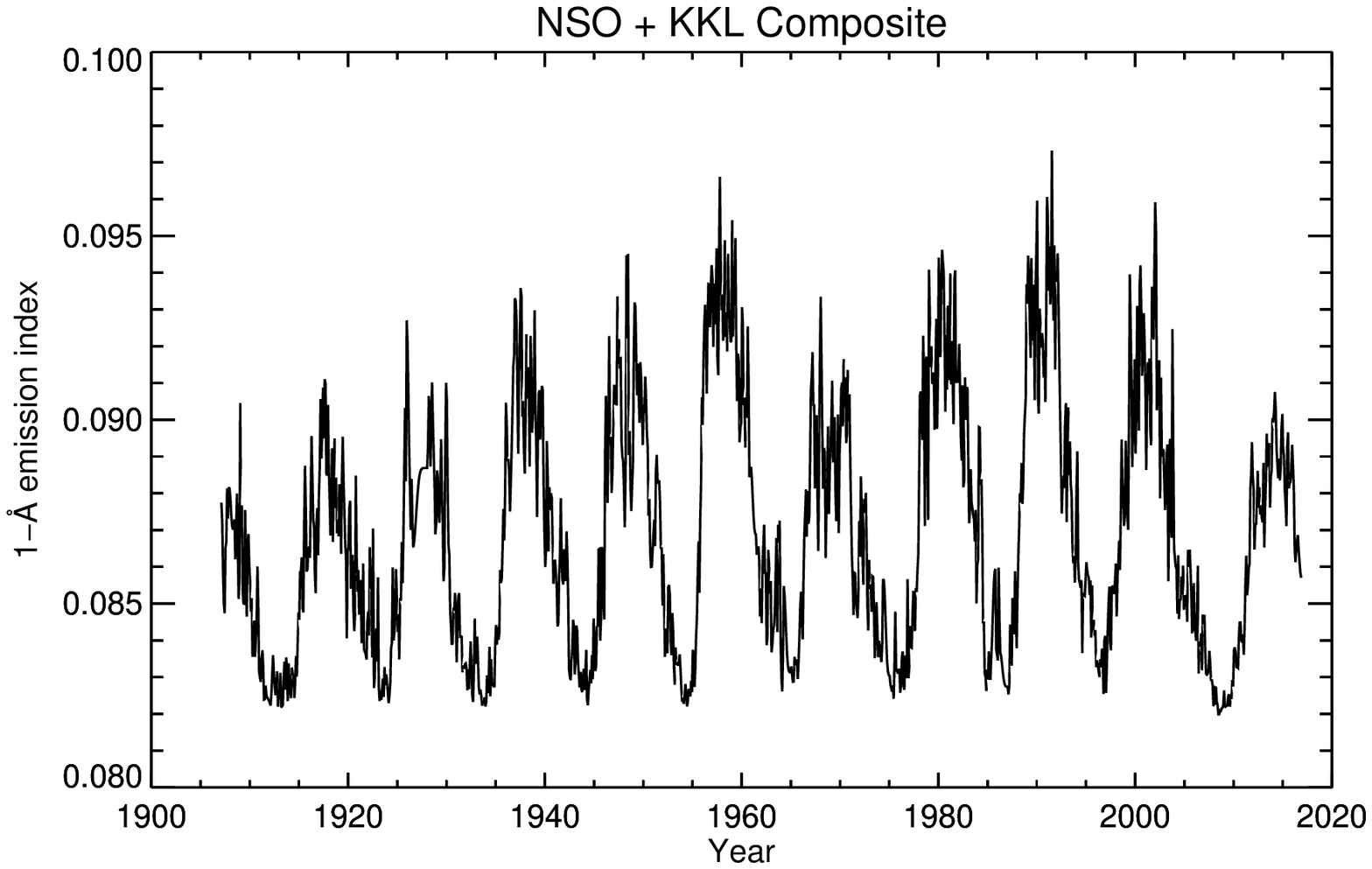}
\caption{Monthly values of the Ca II K 1-\AA~ emission index composite. Data from NSO and Kodaikanal
measurements (KKL) are used.}
\label{composite}
\end{center}
\end{figure*}

Monthly and yearly averages of the Ca II K plage index computed from 
the daily Ca II K full-disk measurements taken at the Kodaikanal Observatory are available since February 1907.
They cover more than 100 years of observations, with a high duty cycle until 1999. Annual mean values
from the NSO composite and the Kodaikanal time series between 1977 and 1999 are used here to derive the coefficients
for the conversion of the Kodaikanal data into the NSO scale. Figure \ref{nso_kkl} shows the excellent correlation between 
the two time series of annual means. 
From the linear fit, we derive the following transformation from plage index to the NSO
Ca II K 1-\AA~ emission index: 

\medskip
KKL$_{\rm NSO}$ = (0.08217 $\pm$ 0.00049) + (0.00020 $\pm$ 0.00001) $\times$ KKL$_{\rm orig}$,
\medskip

\noindent
where KKL$_{\rm orig}$ and KKL$_{\rm NSO}$ are the original Kodaikanal plage index and rescaled emission index values, respectively.
This transformation allows us to combine the two time series together to produce a single composite of the
monthly average 1-\AA~ emission index that goes back to January 1907. Data from the rescaled Kodaikanal series are used
up to the end of 1987. Starting January 1988, the NSO values are used. The final composite is shown in Figure \ref{composite}.

\section{Conclusions}

An overlapping period of observations from December 2006 till October 2015 was used to establish scaling between time series 
of 1-\AA~ emission index from the NSO Sac Peak K-line monitor and SOLIS/ISS. This enables a continuation of  historical 
time series using ISS data. Annually averaged values of 1-\AA~ emission index correlate strongly with other historical 
time series of plage index (area of solar surface covered by the chromospheric plage). 
This allows for the formation of a proxy of 1-\AA~ emission index from 1908 till present.

\bigskip

\noindent
{\bf \large References}

\medskip
Bertello, L.; Pevtsov, A. A.; Harvey, J. W.; Toussaint, R. M. 2011, Solar Physics, Volume 272, Issue 2, pp.229-242.

\medskip
Pevtsov, A. A..; Bertello, L.; Marble, A. R. 2014, Astronomische Nachrichten, Vol.335, Issue 1, p.21.

\end{document}